# Photogenerated singlet oxygen damages cells in optical traps

Stanyslav Zakharov[*] and Nguyen Cong Thanh[**]

*The origin of photodamage of single cells confined in laser tweezers is determined by comparison of action spectra with these recorded for a photoinduced response of cell populations. Triplet-singlet transitions in dissolved oxygen provide a common cause of both phenomena. Such transitions are forbidden in isolated $O_2$ molecules, but occur in solutions due to the formation of temporary oxygen-solvent complexes. In dependence on the photon energy oxygen molecules may be excited alone or in combination with roto-vibrational excitations in $H_2O$. The photodamage is a consequence of a toxic action produced by the singlet oxygen $^1O_2$ ($^1\Delta_g$).*

Optical traps (laser tweezers) combining a cw laser and high numerical aperture microscope, are unique instruments providing the confinement, manipulation, and controlling micro-sized objects by the light pressure [1, 2]. Single cells are the most promising matter for investigations using tweezers [3]. In far red and near infrared commonly employed for the cell trapping, a spectral-selective cell damage, so-called "opticution", was originally observed two decades ago by Ashkin [4] (who developed optical tweezers) and then confirmed by other researchers. To exclude the photodamage it is necessary to reduce the intensity, however, this results in a serious restriction of the technique because the trapping of large or irregularly shaped cells requires higher laser powers. For developments of effective means against the opticution an underlying mechanism should be discovered. Diverse hypotheses were proposed including local heating [5], two-photon absorption [6, 7], intracellular generation of the second harmonic of the incident radiation [8, 9], photodynamic effect mediated by a sensitizer molecule [10-12], production of oxygen-dependent photoreactive species [12]. Despite many experimental efforts the origin of the photodamage remains unclear till now.

---

[*] Lebedev Physical Institute, Russian Academy of Sciences, Moscow.
[**] Institute of Physics, Vietnamese Academy of Science and Technology, Hanoi

A proven way to check up a photobiological hypothesis is measurement of the action spectrum i.e. the spectral dependence for a photoinduced effect. Recording the photoresponse-versus-light intensity relation is desirable, as a linear relation is evidence that the phenomenon is obeyed to the reciprocity law. Then the action spectrum would be similar to the absorption spectrum of a chromophore; its identification can be fulfilled by comparison with spectra of the known molecules. A square-law curve indicates a two-photon process.

Action spectra were recorded in a few studies using predominantly tunable Ti:sapphire lasers [9, 12, 13-15]. Taking together the data suggest that the opticution occurs within the same narrow bands about 760, 900, and (for Nd:YAG lasers)1064 nm, regardless the cell type. A particularly remarkable observation is that photoresponse is linearly, rather than quadratic, related to the laser intensity and is decreased to a background as oxygen is removed from cell environment [12]. Moreover, the level of the photodamage is determined by the time-averaged laser power, not the peak power [16]. It was concluded that unknown mechanism involves a one-photon process mediated by the oxygen [12], and a single parameter—the laser energy—can be used to describe the limitation on cell viability [16].

The following step was taken when the action spectrum for chromosomal disruptions in [13] has been compared with action spectra for other photobiophysical phenomenon named the light-oxygen effect (LOE) [17]. (The LOE is analogous to the photodynamic effect [18] because singlet molecular oxygen is a common reactive species produced, in the latter case, by a sensitization whereas in the LOE by direct excitation. (Note that the first observation of direct excitation of oxygen in a $^1D_g$ state has been made in the gas phase and chlorofluorocarbon solutions at pressure more than 100 atm [19]). The 760 nm band was founded to coincide with one of bands in the LOE action spectrum previously identified as being due to the $(b^1\Sigma_g^+ \leftarrow X^3\Sigma_g^-)$ (0, 0) transition of the molecular oxygen [20]. (In water solutions $O_2$ ($^1\Sigma_g^+$) has not time to reveal its reactivity because of the rapid ($10^{-10}$-$10^{-11}$ s) relaxation to the lowest electronic excitation state $a^1\Delta_g$ possessing a high chemical reactance and commonly called

"singlet oxygen", $^1O_2$). In addition, it was shown that $^1O_2$ generated by photosensitizing can induce chromosome injuring in cellular nuclei [21]. The conclusion has been made that for given spectral portion the dissolved oxygen is the target for photons inducing the cell photodamage in tweezers [17].

Spectral features of other band near 900 nm [14] do not bear any superficial resemblance to the absorption spectrum of oxygen, cellular species or water therefore its origin remained puzzling for long. Recently an approach to unravel the puzzle was proposed [22]. It is based on the concept of simultaneous transitions, i.e. one-photon two-molecule processes induced by intermolecular interactions (for review see [23]). We proposed that the photodamage may be also caused by the singlet oxygen, but generated in a combination with the excitation of solvent (water) molecule(s); a resulting spectral shift masks features of the real reactive species. Such transitions are not exotic; e.g. the singlet oxygen was generated upon irradiation of a solvent - oxygen cooperative absorption band [24]. These can be facilitated by presence of the $O_2/H_2O$ van der Waals complexes, which appear to be found out in liquid water as a result of long-standing studies of a photonucleation mechanism in water vapor [25, 26]. In the case under discussion, the ($a^1\Delta_g \leftarrow X^3\Sigma_g^-$) transition in $O_2$ with the simultaneous excitation of a vibration in $H_2O$ was found to satisfy data from [14], however, we have not been able then to determine what vibration mode is excited because of unavailability of experimental errors in [14]. In ref. [12], which became known to us recently, the tweezer action spectrum within the 790-1064 nm region was recorded more carefully. It has been established that the 900 nm "band" consists, in fact, of two relatively narrow bands with maximums about 870 and 930 nm. It is sufficient to complete the analysis initiated in [17]. Here we demonstrated that toxic action of the singlet oxygen generated by direct excitation is the universal mechanism of spectral-selective photodamage of cells in optical traps.

We measured a photoresponse of different biological samples in some portions of a visible and near IR range (560-1280 nm). LOE action spectra into the spectral range relevant for tweezers are presented in Fig. 1. The oxygen band ($b^1\Sigma_g^+ \leftarrow X^3\Sigma_g^-$) (v = 0, 0) centered at λ = 762 nm manifests itself distinctly both for cell suspensions and protein aqueous solutions. In experiments with cells we were limited on long-wavelength tuning to 850 nm, but increasing the photo-induced effect over a background already has been observed in process of displacement to this limit. One of tweezer action spectra [13] is displayed on the same figure (curve 3). First of all, it is clearly visible the coincidence of position and characteristic details for the 762 nm band in the LOE and tweezer records. For comparison the high-pressure oxygen absorption spectrum is shown in Fig. 2 (to our knowledge, data on the infrared absorption spectrum of oxygen dissolved in water are absent). The assignment of the 762 band to one of bands in the oxygen absorption spectrum ($b^1\Sigma_g^+ \leftarrow X^3\Sigma_g^-$, 0←0 transition) is unequivocally.

Other tweezer action spectra are shown in Fig. 3. Judging from the position of bands in more long wavelength part of the $O_2$ absorption spectrum Fig. 2, it may conclude that the photodamage in tweezers using 1064 nm radiation (Nd:YAG lasers) corresponds to generation of the singlet oxygen due to the ($a^1\Delta_g \leftarrow X^3\Sigma_g^-$) (v = 1, 0) transition. A spectrum in Fig. 3 (curve 1) is important in further analysis because of its connecting-link character. Here both already identified band (760 nm) and unknown "band" about 900 nm are presented. Their amplitudes are approximately equal that is an indirect indication of a like origin of cellular damage. The action spectrum in Fig. 3 (curve 2) does not grasp the 762 nm band, but then two close bands are resolved in the 900 nm portion. We assumed that each of them is associated with generation of singlet oxygen in simultaneous transitions including excitation of vibrational modes in water molecules

$$\{O_2 / H_2O\} \xrightarrow{h\nu} \{O_2^* / H_2O^*\}.$$

Obviously only two oxygen electronic transitions should be considered that vibration energy contributions were positive:

$$O_2: a^1\Delta_g \ (v = 0) \leftarrow X^3\Sigma_g^- \ (v = 0),$$

$$O_2: a^1\Delta_g \ (v = 1) \leftarrow X^3\Sigma_g^- \ (v = 0).$$

This circumstance significantly simplifies the identification problem. An energy price of the first, (0←0) transition as determined from position of a wavelength maximum (1264 nm) in a LOE action spectrum is equal to 7911 cm$^{-1}$ (about 7900 cm$^{-1}$ for a gaseous mixture [28]). The energy for the (1←0) transition obtained from absorption spectra for high-pressure gas and oxygen dissolved in chlorofluorocarbons, is approximately 9390 cm$^{-1}$ (1065 nm) [28]. Spectral shifts for aqua solutions are expected to be not more then a few tens of cm$^{-1}$ [27]. As it is seen from Table 1, an energy difference between positions of the bands in the tweezer action spectrum and oxygen absorption spectrum could be corresponded to water vibrations.

**Table 1.** Vibrational and librational modes of $H_2O$ molecules in liquid phase [29]

| Assignment | Wavelength, μm | Energy, cm$^{-1}$ |
|---|---|---|
| $L_1$, libration | 25 | 396 |
| $L_2$, libration | 15 | 686 |
| $\nu_2$, bend vibrations | 6.08 | 1645 |
| $\nu_2 + L_2$, combination | 4.65 | 2150 |
| $\nu_1$, symmetric stretch | 3.05 | 3277 |
| $\nu_3$, asymmetric stretch | 2.87 | 3490 |

Combination fitting the 870 nm band is simple. This is the (0-0) transition of $O_2$ and H-O asymmetric stretching vibration in $H_2O$ (Table 2) in accordance with an assumption advanced in [22]. For the 930 nm band a similar combination $O_2$ ($^1\Delta_g$ or $^1\Sigma_g^+$) with any single water vibration is not found, therefore two identical $L_2$ librations should be excited to satisfy to necessary conditions. The probability of such an extravagant process seems enough insignificant. More likely possibility is the analogy to the oxygen 919 nm band in the absorption spectrum on Fig. 2. It can be interpreted as the (1-0) transition of $O_2$ with simultaneous excitation of a fundamental vibration mode (about 1500 cm$^{-1}$) in a neighboring molecule of oxygen. Such events connect with existence of short-living $O_2 \cdot O_2$ dimers often named dimols [30]. Dimol bands about 639 and 587 nm, corresponding to electronic transitions in each of members of the oxygen pair were recorded earlier in LOE action spectra [see reviews 27, 31].

**Table 2. Results of fitting**

| $\lambda_{max}$, nm/E, cm$^{-1}$ (experiment) | Oxygen transition; $E_{ox}$, cm$^{-1}$ | $\Delta E$, cm$^{-1}$ | $\lambda_{max}$, nm/E, cm$^{-1}$ (fitting) |
|---|---|---|---|
| 760/13157 | $^1\Sigma_g^+ \leftarrow {^3\Sigma_g^-}$ (0, 0); 13123 | 24 | 762/13123 |
| 870/11 363 | $^1\Delta_g \leftarrow {^3\Sigma_g^-}$ (0, 0); 7911 | 3490 | 877/11 400[1] |
| 930/10 750 | $^1\Delta_g \leftarrow {^3\Sigma_g^-}$ (1, 0 ); 9407 | 1343 | 925-935[2] |
| 1064/9398 | $^1\Delta_g \leftarrow {^3\Sigma_g^-}$ (1, 0); 9407 | - 9 | 1065/9390 |

[1] + H$_2$O asymmetric stretch vibration; [2] + fundamental vibration in O$_2$.

In conclusion, the origin of the photodamage (opticution) has been elucidated in all spectral range where tweezers operate. In accordance with preliminary findings it is an oxygen-dependent, cell type-independent one-photon process determined by wavelength and energy of laser beam. Unlike the previous expectations, dissolved oxygen is a primary acceptor of photons rather than any biomolecules or sensitizers whereas reactive species involved is the singlet oxygen. As the quantity of the singlet oxygen generated is linearly related to the product of the laser energy by oxygen concentration, this single parameter ("photo-oxygen load") can be used to describe the limitation on the photodamage. The knowledge of the underlying mechanism should prove useful not only for a well-founded choice of the light wavelength for optical trapping. Lasers are widely applied in biophysical investigations and medical diagnostic techniques such as multiphoton and fluorescent microscopy or flow cytometry. When these are used in an interaction with living samples the main concern is with exception of side thermal effects. From the presented analysis it is evident the necessity to pay not smaller attention to check the absence of singlet oxygen's damaging action. Molecular oxygen has many inherent and, very likely, water-combined absorption bands in ultraviolet, visible and near infra-red ranges. The majority of them are not studied till now. Therefore caution should be exercised in long-term cell irradiation even at small light fluxes.

**Acknowledgments**

We thank Boris Dgagarov for the oxygen absorption spectrum provision, and Alexey Kryukov for help at the manuscript preparing. This work was partially supported by the Russian Foundation for Basic Research under Project No. 09-02-90300-Viet p.

**Fig 1.** Action spectra (normalized to 1) for the LOE photo-activation of erythrocyte suspensions (curve 1) [27], albumin aqua solutions (curve 2) [27], and tweezer photodamage of chromosomes in single rat kangaroo (curve 3) [13] (with permission).

**Fig 2.** Absorption spectrum of high-pressure molecular oxygen.

**Fig 3.** Action spectra (normalized to1) for tweezer photodamage in single Chinese hamster ovary cells (curve 1) [14] and in E. coli cells (curve 2) [12] (with permission).

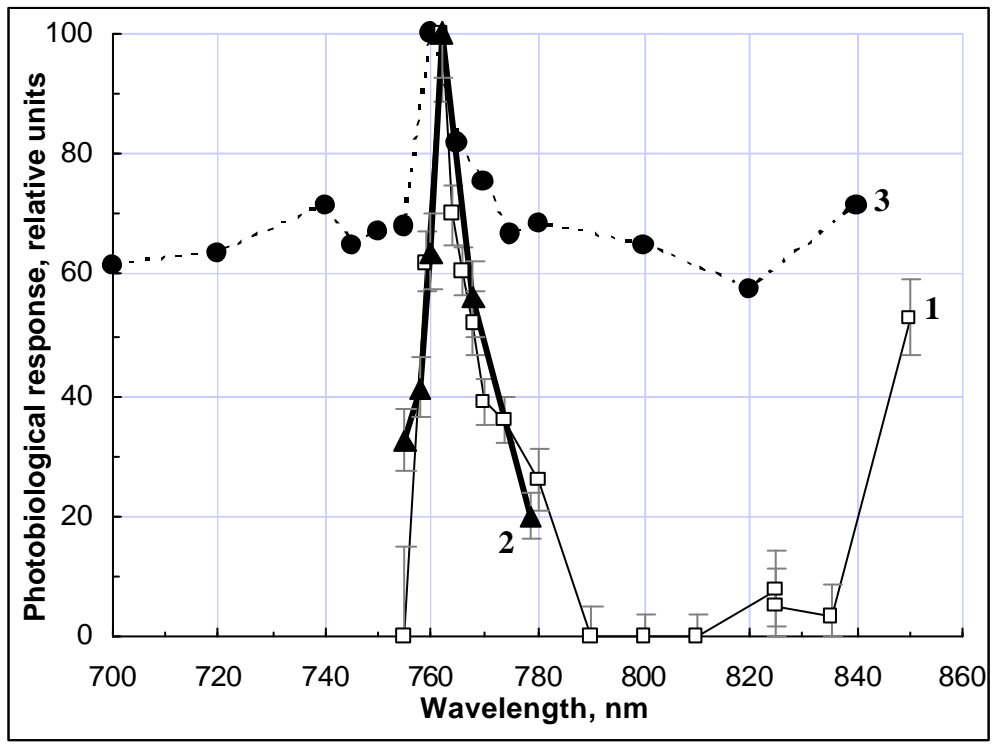

**Fig 1**

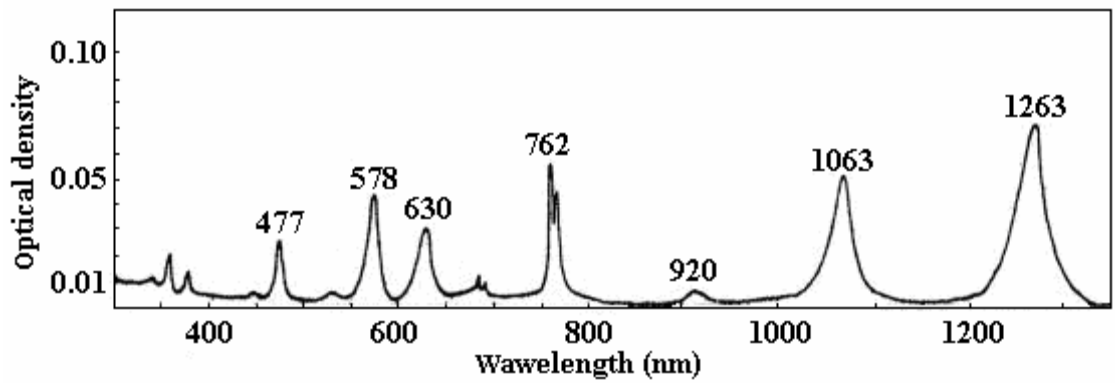

**Fig. 2**

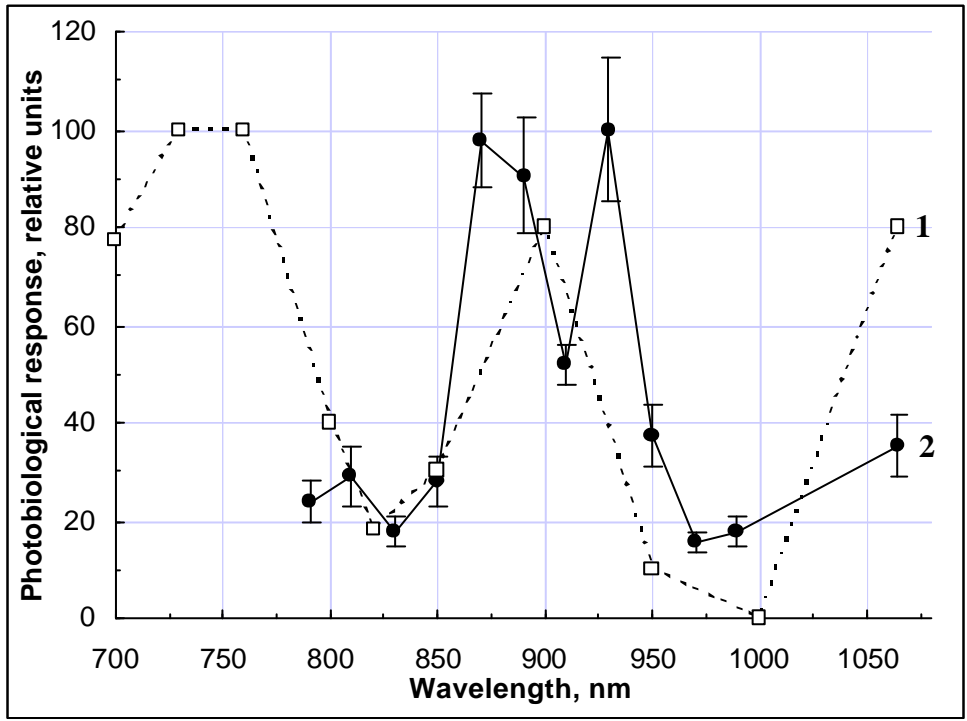

**Fig. 3**